# A Comprehensive Review on Power System Risk-Based Transient Stability

Umair Shahzad
Department of Electrical and Computer Engineering,
University of Nebraska-Lincoln,
Lincoln, NE, USA
umair.shahzad@huskers.unl.edu

*Abstract*–Power systems are getting more complex than ever and are consequently operating close to their limit of stability. Moreover, with the increasing demand of renewable wind generation, and the requirement to maintain a secure power system, the importance of transient stability cannot be overestimated. Considering its significance in power system security, it is important to suggest a different methodology for enhancing the transient stability, considering uncertainties. Current deterministic industry practices of transient stability assessment ignore the probabilistic nature of variables (fault type, fault location, fault clearing time, etc.). These approaches typically provide a cautious principle and can result in high-priced expansion projects or operational limits. With the increasing system uncertainties and widespread electricity market deregulation, there is a strong inevitability to incorporate risk in the traditional transient stability analysis. Accurate assessment of transient stability in a modern power network is becoming a strict requirement both in planning and in real-time operation, due to the increasingly intricate dynamics of a power system. Further, increasing sources of uncertainty in forecast state and in the reaction to faults highly implies the implementation of risk-based approach in assessing transient stability. Thus, this paper aims to provide a comprehensive review of risk-based transient stability in power networks and the accompanying research. It is believed that this review can be an inception for researchers in the domain of power system planning and security.

I. INTRODUCTION

Driven by various techno-economic and environmental factors, the electric energy industry is anticipated to undergo a paradigm shift, with a significantly augmented level of renewables, especially, wind and solar power sources, gradually replacing conventional power production sources (coal, diesel, natural gas, etc.) [1-2]. This increasing demand of large-scale wind integration in the conventional power system, along with the inherent and external uncertainties of the system, brings a lot of challenges [3-4]. Power systems are regularly exposed to unanticipated faults. Such faults can cause transient instability and can consequently lead to prevalent outages [5]. With the implementation of a deterministic criterion for system stability, power systems generally operate with a large stability margin. Usually, these deterministic criteria provide safe, but conservative limits for system operating conditions. The most critical security criterion is the (*N*-1) security criterion that guarantees safe operation of the power system, after the failure of a single element of the system, where *N* is the total number of system components [5].

In recent times, various sources of probabilistic renewable energy generation are on the rise. These uncertainties, coupled with load uncertainties, are becoming the key features of contemporary power networks [6]. The current industry practices use the deterministic approach for Transient Stability Assessment (TSA) [7-8]. Although, the deterministic approaches result in highly stable power systems, but they fail to incorporate the probability of various system components and conditions. In addition to the great expense due to conventional models, the key drawback with the deterministic assessment techniques is that they consider all stability problems to have equal risk [9]. Various literature [8-14] mention that probabilistic risk-based transient stability (RBTS), and incorporating risk in power planning procedures, is a future research area, and consequently, work needs to be done in this domain. Moreover, planning manuals of various utilities [15-18] recommend using risk-based probabilistic approaches in the near future. Moreover, the integration of renewable generation, introduces more stochasticity in the system, making the application of probabilistic practices essential in the TSA process.

The rest of the paper is organized as follows. Section II discussed the background of deterministic and probabilistic transient stability (PTS). Section III presents literature review regarding RBTS. Section IV discusses the research gaps and recommendations for future work. Finally, Section V concludes the paper.

II. DETERMINISTIC AND PROBABILISTIC TRANSIENT STABILITY

Conventionally, deterministic criterion has been used for transient stability evaluation for power system planning and operation [19-20]. This method is generally considered for a single operating condition, commonly known as the worst-case scenario. In most cases, the (*N*-1) contingency principle is used. This means that individual system components are removed one by one for the analysis. The worst-case scenario then gets transformed to numerous extreme operating conditions, together with several most critical contingencies, for which the system should be designed to withstand. Although this worst-case approach has served the industry well; however, in a deregulated environment, the utilities require to know the risk value [21]. The deterministic approach has at least the following three drawbacks [22-23].

*1)* "Only consequences of contingencies are evaluated, but probabilities of occurrence of contingencies are ignored. Even if the consequence of a selected contingency is not very severe, system risk could still be high, if its probability is relatively large. Conversely, if the probability of an outage event is extremely small, the contingency analysis of such an event may result in an uneconomic operational decision.

*2)* All uncertain factors that exist in real life (such as uncertainty of load variations, variability of renewable generation, random failures of system components, fuzzy factors in parameters or input data, errors in real-time information, volatility of power demand on the market, etc.) are ignored in the deterministic analysis. This can lead to results, biased from the reality.

*3)* The deterministic approach is based on pre-selected worst cases. In implementation, however, the actual worst case may be missed [23]."

Moreover, as the result of deterministic stability analysis is binary (stable or unstable), therefore, the transient instability risk cannot be quantified. Therefore, examining the system transient stability by applying risk assessment has become a critical research technique [24]. A pictorial representation of a standard framework for deterministic TSA is shown in Fig. 1.

With the current drift towards competitive and deregulated electricity market environment, the power utilities are required to guarantee, besides a safe reliability level, an economical operational efficiency. In these situations, a probabilistic assessment approach becomes tremendously beneficial [25]. "The probabilistic studies consider the stochastic and probabilistic nature of the real power system. It considers the probability distribution of one or more uncertain parameters, and hence, reflects the actual system in a better manner. Although, it has been long established that deterministic studies may not sufficiently characterize the full extent of system dynamic behavior, the probabilistic approach has not been extensively used in the past in power system studies, mainly due to lack of data, limitation of computational resources, and mixed response from power utilities and planners [20, 22-23]. Probabilistic approaches are mainly appropriate, for the examination of a system, with randomness and uncertainty, which are obviously the main features of future power networks."

In the past several years, there has been a considerable increase in connections of intermittent and stochastic, power electronics interfaced renewable energy generation sources. These uncertainties, coupled with load uncertainties, are becoming one of the crucial characteristics of modern power systems. The TSA of such systems using traditional deterministic methodology is swiftly becoming inappropriate and thus, unique probabilistic assessment methods are desirable, and are being established [26].

Although, the deterministic approaches result in very secure power systems, but they blatantly ignore the stochastic characteristics of a real power network. Also, with the introduction and acceptance of competitive electricity markets and intricacy in solving practical problems of power system planning, the deterministic methods are insufficient and obsolete [27]. This, along with the rising power system, uncertainties have greatly motivated the application of probabilistic methodologies, for TSA. A pictorial representation of a typical framework for probabilistic TSA is shown in Fig. 2.

To the best of author's knowledge, there exists no work which comprehensively reviews RBTS of power systems. Thus, the main objective and contribution of the current paper is to review work related to RBTS and provide research gaps and recommendations for future work.

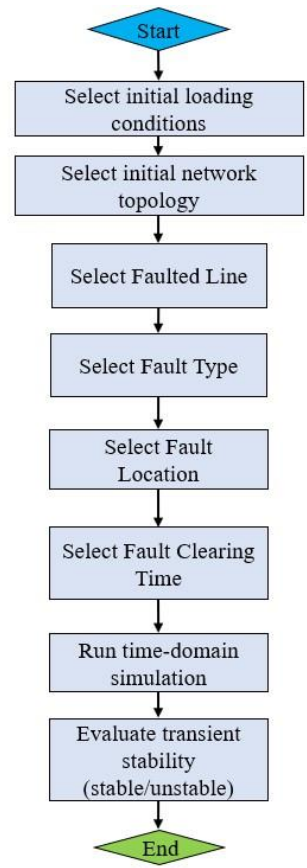

Fig.1. Framework for deterministic TSA

TABLE I. RISK VALUES FOR TWO DIFFERENT CONTINGENCIES

| Contingency | Probability | Impact | Risk |
|---|---|---|---|
| *C1* | 0.05 | 20 | 1 |
| *C2* | 0.02 | 30 | 0.6 |

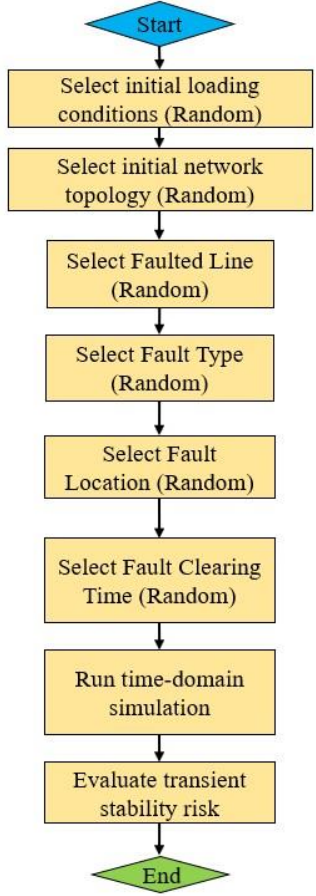

Fig. 2. Framework for probabilistic TSA

### III. LITERATURE REVIEW: RISK-BASED TRANSIENT STABILITY

The product of probability of an unforeseen event and its impact is commonly known as risk, which is generally mathematically defined as (1) [28-32].

$$Risk = \sum_i \Pr(E_i) \times Sev(E_i) \qquad (1)$$

where $E_i$ is the ith event (contingency) and $\Pr(E_i)$ is its probability. $Sev(E_i)$ quantifies the impact of $E_i$.

The risk is the system's exposure to failure and is generally determined by considering both the probability of occurrence of an event and the impact of the event. The deterministic stability assessment introduces operating limits, based on the impact of contingencies. However, based on the risk-based approach, these operating limits are calculated by using the weighted sum of risk components of all the contingencies, considering both the probability and the impact [29]. A simple example can be used to outline the significance of using risk in power systems. Consider two contingencies (*C1* and *C2*), along with their probability of occurrence and the corresponding severity (impact), as outlined in Table I. If decision-making is assumed to be based on deterministic criteria, *C2* is found to be more severe as its impact is greater than *C1*; however, if risk-based (considers both probability and impact) decision-making is used, the converse is true.

Risk-based approach describes possibility of contingency by probability, and the corresponding impact (or consequence) by severity function. The product of this probability and associated severity is termed as risk. In risk-based stability assessment, the risk index consists of each possible contingency occurrence probability and the associated impact [33]. The first attempt toward RBTS was proposed in [34] and [35], where the notion of limiting operating point functions was used. These functions return the limiting generation level for any fault type and fault location. Reference [36] used risk-based approach, to analyze the transient stability of power networks, incorporating wind farms. The proposed methodology of transient instability risk assessment is based on the MC method and eventually, an inclusive risk indicator, based on angle and voltage stability, is devised. The work considered only three phase line faults.

Reference [37] presented a distributed computing approach for transient stability analysis, in terms of measuring critical clearing time and the overall risk index, for various uncertainties. The work considered only a three phase to ground fault. [38] presented a method to determine the risk of transient stability. It described the application of rotor trajectory index (RTI), to assess the severity of power systems, when it was subjected to a three-phase fault. The RTI was suggested as an index used to represent severity of transient instability. Reference [39] focused on risk of transient instability. A procedure was suggested to evaluate the potential loss of synchronism of a generator, in terms of probability and consequences. A transient risk assessment method, based on trajectory sensitivity, was presented in [40].

In [41], transient stability risk assessment framework incorporating circuit breaker failure and severe weather was presented. All related random variables, such as load demand, fault type, fault location, and fault clearing time (FCT) were considered using appropriate probability density functions (PDFs). Reference [42] proposed a two-stage power system TSA method based on snapshot ensemble long short-term memory (LSTM) network to quantify the risk of transient stability. The two stages consisted of dynamic hierarchical assessment and application of regression to screen credible samples and predict their transient stability margin, respectively.

Reference [43] provided an algorithm for the rapid on-line transient instability risk assessment related with existing or forecasted operation conditions, using IEEE 39-bus system. The transient instability risk was defined in terms of the probability of transient instability and its cost. [44] used Radial Basis Function Neural Network (RBFNN) to compute transient stability risk in a cyber physical power system. It used a synchro phasor data analytics based predictive algorithm for detecting vulnerable generating units transgressing the stability boundaries for proactive control actions. Some other research work associated with RBTS can be found in [45-47].

## IV. RESEARCH GAPS AND FUTURE RECOMMENDATIONS

From the extensive literature review of major works in RBTS, the following gaps were identified. The major research gap in these works include not considering all the fault events (faulted line, fault type, fault location, FCT) randomly, i.e., only some of the events are considered random variables, while others are considered as deterministic. To fully encapsulate the concept of risk, it is important to consider all parameters. Moreover, there is a string need to establish a standard risk metric for transient stability. Majority of research takes a very simplistic approach to treat the randomness of associated variables (load, renewable generation, component availability, fault type, generator fuel price, etc.). Also, there is a need to extensively include renewable generation and external weather events while assessing this risk. There is a need to establish advanced softwares and tools which can deal with probabilistic analysis of large-scale power systems to ensure accurate RBTS assessment. It is important to consider more than single contingency as ignoring the higher order contingences will result in a very conservative analysis of transient stability. Big data analytics must be researched further to delve into the various possibilities of data regarding random input and output variables of power system, and to comprehend unknown patterns, correlations, market developments and customer preferences.

Given the exceptional changes in the electric power industry, and the stringent requirements to maintain system reliability at a minimum cost, RBTS planning is becoming more complex than ever. In this regard, some significant recommendations in this are outlined below [13]. All the existing North American Electric Reliability Corporation (NERC) transmission planning standards are deterministic. However, recently NERC has shown interest in considering probabilistic approaches in transmission planning and organized multiple workshops in Eastern Interconnection as well as in Western Electricity Coordinating Council (WECC) on risk-based planning. NERC and states, in collaboration with other stakeholders can collaborate, and develop a long-term vision for establishing risk-based planning framework. This is an important domain which requires noteworthy effort and coordination.

As mentioned before, risk-based planning requires active research and industry participation for its wider adoption. States can encourage research efforts and work closely with research organizations, universities, national labs, commercial software developers, and utility industry to ensure that research needs are addressed, and practical solutions are proposed. The research community needs to work more closely with the industry to clearly demonstrate the benefits of probabilistic methods. The industry needs to clearly communicate inadequacies in deterministic methods and areas that probabilistic methods can be most productive.

This study provided a review of some major research works in RBTS. This can be a remarkable starting point for researchers in the domain of power system stability and security. Recent research [48-53] reveals that there is a lot of work which needs to be done in the domain of RBTS of modern power systems.

## V. CONCLUSION AND FUTURE WORK

Power system transient stability is an integral part of power system planning and operation. Traditionally, it has been assessed using deterministic approach. Also, current North American Electric Reliability Corporation (NERC) reliability standards are deterministic and do not include any probabilistic methods. With the increasing system uncertainties, environmental pressures of incorporating green energy, and widespread electricity market liberalization (deregulation), there is a strong need to incorporate risk in conventional transient stability analysis in transient stability evaluation. RBTS can entirely consider both the probability and impact of the unexpected event (fault), and thus, can quantify the indicator of risk. This method has a greater logic compared to conventional methods as uncertainty, renewable generation, and deregulation are characteristics of modern power system.

This paper provided the foundation for more comprehensive research for RBTS assessment in power system. Research on power system RBTS is just the tip of iceberg. Unknown events will always be a formidable challenge to power system researchers, and can lead to large values of risk, which eventually can result in social, economic, and technical losses. Thus, as a future work, it is important to research and formulate reasonable methods to assess this risk. It is believed that this review would provide a good offset for any future research in the domain of power system planning, particularly, power system security and stability. A novel and unique new power system model, incorporating both network topology changes and (*N*-2) contingency criterion, will provide an excellent practical kick-off for RBTS of modern power systems.